\documentclass[aps,pra,twocolumn,showpacs,preprintnumbers,amsmath,amssymb]{revtex4}
\usepackage{graphicx}
\usepackage{bm}
\usepackage{dcolumn}

\def\prv#1#2#3{Phys. Rev. {\bf #1}, #2 (#3)}
\def\rmp#1#2#3{Rev. Mod. Phys. {\bf #1}, #2 (#3)}
\def\prl#1#2#3{Phys. Rev. Lett. {\bf #1}, #2 (#3)}

\def\prb#1#2#3{Phys. Rev. B {\bf #1}, #2 (#3)}

\def\pre#1#2#3{Phys. Rev. E {\bf #1}, #2 (#3)}

\def\epjb#1#2#3{Eur. Phys. J. B {\bf #1}, #2 (#3)}

\def\ejp#1#2#3{Eur. J. Phys. {\bf #1}, #2 (#3)}

\def\zp#1#2#3{Z. Phys. {\bf #1}, #2 (#3)}

\def\noi{\noindent}
\def\bc{\begin{center}}
\def\ec{\end{center}}
\newcommand{\bea}{\begin{equation}}
\newcommand{\eea}{\end{equation}\noi}
\newcommand{\ber}{\begin{eqnarray}}
\newcommand{\eer}{\end{eqnarray}\noi}
\begin{document}
\title{A complete theory for the magnetism of an ideal gas of electrons}
\author{Shyamal Biswas$^1$}\email{sbiswas.phys.cu@gmail.com}
\author{Swati Sen$^2$}
\author{Debnarayan Jana$^1$}
\affiliation{$^1$Department of Physics, University of Calcutta, 92 APC Road, Kolkata-700009, India\\
$^2$Department of Physical Sciences, Indian Institute of Science Education \& Research-Kolkata, Mohanpur-741252, India}
\date{\today}

\begin{abstract}
We have explored Pauli paramagnetism, Landau diamagnetism and de Haas-van Alphen effect in a single framework, and unified these three effects for all temperatures as well as for all strengths of magnetic field. Our result goes beyond Pauli-Landau result on the magnetism of the 3-D ideal gas of electrons, and is able to describe crossover of the de Haas-van Alphen oscillation to the saturation of magnetization. We also have obtained a novel asymptotic series expansion for the low temperature properties of the system.
\end{abstract}
\maketitle
\section{Introduction}
Pauli paramagnetism \cite{pauli}, Landau diamagnetism \cite{landau-d} and de Haas-van Alphen effect \cite{dhva} are very common topics of condensed matter and statistical physics not only for the undergraduate and graduate students \cite{huang,bhattacharjee,landau-lifshitz,landau-lifshitz2} but also for the theoreticians \cite{steinberg,lai,kita,mineev,lohneysen,lukyanchuk,hussein} and experimentalists \cite{anderson,cooper,singh,shaykhutdinov,gasparov,nascimbene,bergk,ketterle}. The simplest system of interest, for these particular topics, is a 3-D ideal gas of electrons exposed in a constant magnetic field ($\textbf{B}=B\hat{k}$) \cite{huang,landau-lifshitz}. For the existence of spin, each electron behaves as a tiny magnet; and being the value of spin to be $1/2$, the gas of electrons obeys Fermi-Dirac statistics. Response of the external magnetic field ($B\hat{k}$) to the tiny magnets (electrons) of dipole moment $\mu_B\hat{\sigma}_z$, is known as Pauli paramagnetism. On the other hand, magnetic field induces orbital motions to the charged electrons, and closed electric circuits in the $x-y$ plane are produced every where in the space obeying Fermi-Dirac statistics. These circuits, according to the Lenz's law, oppose the external magnetic flux to pass through them, and result Landau diamagnetism. For the case of strong magnetic field, apart from Pauli paramagnetism and Landau diamagnetism, oscillations of magnetization of the free electron gas with period $\sim1/B$ are observed \cite{dhva}. Such a phenomenon is called de Haas-van Alphen effect. These are the inclusive scenario for the magnetization of a 3-D ideal gas of electrons, and are very important in statistical and condensed matter physics, in particular, for the characterization of metals.

\begin{figure}
\includegraphics{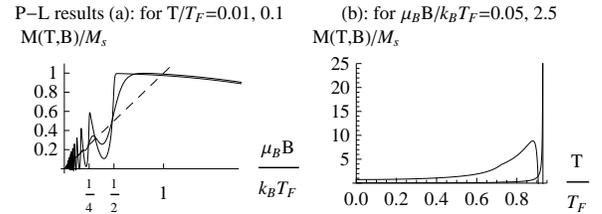}
\caption{Thick and thin solid lines in (a) represent Pauli-Landau (P-L) results on magnetization of the ideal gas of electrons for $T/T_F=0.01$ and $0.1$ respectively. Dashed line represents weak field result for $T\rightarrow0$. Thick and thin solid lines in (b) represent P-L results for $\mu_BB/k_BT_F=2.5$ and $0.05$ respectively.}
\end{figure}

That electrons have nonzero magnetic moment for their intrinsic spin ($1/2$), and that they obey Fermi-Dirac statistics, was experimentally verified after Pauli's prediction of the paramagnetism of the electron gas in a metal exposed in a weak magnetic field \cite{pauli}. Discrepancy of Pauli's prediction and experimental data \cite{schumacher} for weak magnetic field was attributed by Landau's prediction on the diamagnetic contribution from orbital motions of the electrons \cite{landau-d}. Landau showed that, this diamagnetic contribution to the lowest order in magnetic field, is $1/3$ of the paramagnetic contribution predicted by Pauli. Thus, in presence of a weak magnetic field, ideal gas of electrons is a paramagnet. Although it is a common practice to consider magnetization of the electron gas as a separate sum of Pauli's and Landau's results \cite{huang}, Landau unified Pauli paramagnetism and Landau diamagnetism in a single framework to justify the separate addition of the two contributions \cite{landau-lifshitz,landau}. But, this unification was made only to the lowest order in magnetic field. For strong magnetic field, Landau obtained a formula for the oscillatory part of the magnetization, and predicted it to be the theory for the de Haas-van Alphen effect \cite{landau}. But, he did not unify the three effects altogether except to the second lowest order in magnetic field. It was a natural question, whether the $1/3$ relationship valid for all strengths of the magnetic field. If not, how does it change for all values of the magnetic field? This question was partially answered by Sondheimer-Wilson \cite{sondheimer,wilson,molinari}. They unified the three effects for weak as well as for strong magnetic field, and obtained the magnetization as a sum of Pauli's weak field formula for the paramagnetism, Landau's weak field formula for the diamagnetism, and Landau's strong field formula for the de Haas-van Alphen effect. Although essentially no new result, after Landau's calculation \cite{landau}, was obtained by Sondheimer-Wilson, yet they indicated a way to go beyond Pauli-Landau theory for the three effects in particular for the low temperature regime \cite{wilson}. However, Pauli-Landau result on the 3-D ideal gas of electrons can also be represented by Sondheimer-Wilson formula \cite{landau,sondheimer,wilson,molinari}.

Being the topic old, before going into the details, we should explain our motivation of writing this paper. Let us start from Pauli-Landau result (or Sondheimer-Wilson formula) as presented in FIG. 1. Incompleteness of Pauli-Landau result on the magnetism of the 3-D ideal gas of electrons is clearly evident in this figure. It neither explains the saturation of magnetization of the ideal gas of electron nor it does have a classical limit. So, Pauli-Landau (or Sondheimer-Wilson) theory needs to be completed for all strengths of magnetic field as well as for all temperatures. In the following, we will complete their theory analytically for all temperatures and fields.

Calculations of this paper will begin with a generalized form of the Landau levels for the 3-D ideal gas of electrons exposed in a constant magnetic field. Then we will proceed adopting the steps of Landau-Lifshitz for unifying Pauli paramagnetism and Landau diamagnetism and for the quantitative introduction of de Haas-van Alphen effect \cite{landau-lifshitz}. Then we will proceed in our own way to unify these three effects and to get the most general form of the grand potential of the system. Therefrom we will evaluate magnetization, susceptibility and thermodynamic energy. We will plot field and temperature dependence of these thermodynamic quantities, and compare our results with that of Pauli-Landau \cite{landau,landau-lifshitz}. For the low temperature regime, we will also generalize Pauli-Landau (or Sondheimer-Wilson) formula in a novel way \cite{landau,landau-lifshitz,wilson}.

\section{Grand potential}
Let us consider a system of 3-D ideal gas of electrons in equilibrium with a heat and particle reservoir of temperature $T$ and chemical potential $\mu$. Let the total average number of particles (electrons) be $N$, and the mass of each particle be $m$. Let also the electron gas be exposed in a constant magnetic field $\textbf{B}=B\hat{k}$ along the $z$ direction. Now, the energy levels of each particle of the system are given by the generalized form of the Landau levels \cite{landau-lifshitz}
\begin{eqnarray}\label{eqn:1}
\epsilon_{j,p_z}=\frac{p_z^2}{2m}+2\mu_BBj
\end{eqnarray}
where $\textbf{p}_z$ is the momentum of a particle along the $z$ direction, $\mu_B$ is the Bohr magneton, and $j$ represents the $j$th level which has a degeneracy \cite{landau-lifshitz}
\bea\label{eqn:2}
g_j=\bigg\{\begin{matrix}&1 \ \ \ \ \ \ \ \ \ \ \ \ \ \ \ \ \ \ \ \ \ \ \ \ \ \ \ \ \ \ \ \ \ \ \ \ \ \ \ \ \ \text{for} \ \ j=0\\
&2 \ \ \ \ \ \ \ \ \ \ \ \ \ \ \ \ \ \ \ \ \ \ \ \ \ \ \ \ \ \ \ \text{for} \ \ j=1, 2, 3,... .
\end{matrix}
\eea

\subsection{Landau-Lifshitz integral form of grand potential}
Grand potential of our system can be obtained from Eqns.(\ref{eqn:1}) and (\ref{eqn:2}) as \cite{landau-lifshitz}
\begin{eqnarray}\label{eqn:3}
\Omega=2\mu_BB\bigg[\frac{1}{2}f(\mu)+\sum_{j=1}^\infty f(\mu-2\mu_BBj)\bigg],
\end{eqnarray}
where $f(\mu)$ is given by \cite{landau-lifshitz}
\begin{eqnarray}\label{eqn:4}
f(\mu)=-\frac{k_BTmV}{2\pi^2\hbar^3}\int_{-\infty}^{\infty}\ln\big[1+\text{e}^{\frac{\mu-p_z^2/2m}{k_BT}}\big]\text{d}p_z,
\end{eqnarray}
$V=AL$ is volume of the system, $A$ is area along $x-y$ plane of the system, and $L$ is extent of the system along the $z$ direction. Summation in Eqn.(\ref{eqn:3}) can be exactly obtained using Poisson summation formula: $\sum_{j=-\infty}^{\infty}\delta(x-j)=\sum_{k=-\infty}^{\infty}\text{e}^{2\pi i kx}$ which can be recast multiplying the both sides by $g(x)$ and integrating over $x$ from $0$ to $\infty$ as
\begin{eqnarray}\label{eqn:5}
\frac{1}{2}g(0)+\sum_{j=1}^\infty g(j)&=&\int_0^\infty g(x)\text{d}x+2\sum_{k=1}^\infty\int_0^\infty\cos(2\pi k x)\nonumber\\&&\times g(x)\text{d}x.
\end{eqnarray}
Substituting $g(j)$ as $f(\mu-2\mu_BBj)$ and putting right hand side of Eqn.(\ref{eqn:5}) into the square bracket in Eqn.(\ref{eqn:3}), we get
\begin{eqnarray}\label{eqn:6}
\Omega&=&2\mu_BB\bigg[\int_0^\infty f(\mu-2\mu_BBx)\text{d}x\nonumber\\&&+2\sum_{k=1}^\infty\int_0^\infty\cos(2\pi k x)f(\mu-2\mu_BBx)\text{d}x\bigg].
\end{eqnarray}
The first term ($\Omega_0$) in the square bracket in Eqn.(\ref{eqn:6}) is independent of magnetic field, as because, under suitable variable transformation: $\mu-2\mu_BBx=y$, we can eliminate $B$ from $\Omega_0=2\mu_BB\int_0^\infty f(\mu-2\mu_BBx)\text{d}x$ to get $\Omega_0=\int_{-\infty}^\mu f(y)\text{d}y$. Now, expanding the logarithm in $f(y)$ and integrating over $p_z$ and $y$, we recast $\Omega_0$ as
\begin{eqnarray}\label{eqn:7}
\Omega_0=\int_{-\infty}^\mu f(y)\text{d}y=-2k_BT\frac{V}{\lambda_T^3}\big[-\text{Li}_{5/2}(-z)\big],
\end{eqnarray}
where $\lambda_T=\sqrt{2\pi\hbar^2/mk_BT}$ is the thermal de Broglie wavelength, $z=\text{e}^{\mu/k_BT}$ is the fugacity, and $\text{Li}_{\sigma}(\textbf{x})=\text{x}+\text{x}^2/2^\sigma+\text{x}^3/3^\sigma+...$ is a polylog function of order $\sigma=5/2$. We already have mentioned that, $\mu$ is the chemical potential of the system in absence of the magnetic field. Thus, for a fixed average number ($N$) of particles (electrons), $\mu$ is to be obtained using the implicit formula \cite{biswas}
\begin{eqnarray}\label{eqn:8}
N=-\frac{\partial\Omega_0}{\partial\mu}=2\frac{V}{\lambda_T^3}\big[-\text{Li}_{3/2}(-z)\big].
\end{eqnarray}
So far we have adopted the steps shown in the celebrated book of Landau and Lifshitz \cite{landau-lifshitz}. Let us now proceed in our own way.

\subsection{Evaluation of integrals in the grand potential}
The summation term in the square bracket in Eqn.(\ref{eqn:6}), on the other hand, can be evaluated expanding the logarithm in $f(\mu-2\mu_BBx)$ and integrating over $p_z$ and $x$ (using the formula $\int_0^\infty\text{e}^{-ax}\cos(bx)\text{d}x=\frac{a}{a^2+b^2}$) to yield
\begin{eqnarray}\label{eqn:9}
\delta\Omega&=&\Omega-\Omega_0=\frac{2\Omega_0}{-\text{Li}_{\frac{5}{2}}(-z)}\sum_{k,j=1}^\infty\frac{(-1)^{j+1}z^j}{j^{5/2}\big[1+\frac{\pi^2k^2}{j^2}(\frac{k_BT}{\mu_BB})^2\big]}\nonumber\\&=&-\frac{Nk_BT}{-\text{Li}_{3/2}(-z)}\sum_{j=1}^\infty\frac{(-1)^{j+1}z^j}{j^{5/2}}\big[jb\coth(jb)-1\big],
\end{eqnarray}
where $b=\frac{\mu_BB}{k_BT}$ is a scaled magnetic field. The second line of this equation is obtained inserting the constraint in Eqn.(\ref{eqn:8}) and summing over $k$ using the Mittag-Leffer expansion of $\coth(x)$: $\frac{1}{x}+2x\sum_{k=1}^\infty\frac{1}{k^2\pi^2+x^2}$. Since no approximation has been made so far, Eqn.(\ref{eqn:9}) represents the most general formula for magnetism of the system of our interest, and of course, is applicable for all possible values (strengths) of the magnetic field as well as for all temperatures. This equation must be the general formula for the Landau diamagnetism, Pauli paramagnetism and de Haas-van Alphen effect. An essentially similar equation, like this, was also obtained by Sondheimer and Wilson within a different (density matrix) formalism \cite{sondheimer}. But, this equation is difficult to work with for all temperatures as well as for all strengths of the magnetic field. While here-from they proceeded for low temperature calculation \cite{wilson}, we will proceed for all temperatures and fields.

For the case of weak magnetic field ($b\ll1$), Laurent series expansion formula of $\coth(jb)$: $\frac{1}{jb}+\frac{jb}{3}-\frac{(jb)^3}{45}+...+\frac{[B_{2k}(2jb)]^{2k}}{(2k)!jb}+...$ is frequently used, and Eqn.(\ref{eqn:9}) is usually approximated within the lowest order in $b$ as \cite{landau-lifshitz}
\begin{eqnarray}\label{eqn:10}
\delta\Omega\approx-\frac{Nk_BT}{3}\bigg(\frac{\mu_BB}{k_BT}\bigg)^2\frac{\text{Li}_{1/2}(-z)}{\text{Li}_{3/2}(-z)}.
\end{eqnarray}
This approximate formula represents the magnetism of our system in weak field case, and it unifies Landau diamagnetism and Pauli paramagnetism \cite{landau-lifshitz,wilson,huang}. This approximate formula, however, is not conclusive from Eqn.(\ref{eqn:9}) as because, the approximation $bj\ll1$ is applicable only for lower values of $j$, but not for the higher values ($\gtrsim\frac{1}{b}$). Besides these, contributions of higher values of $j$ are probabilistically negligible for negative values of the chemical potential, but not for the positive values. Thus, for $b\ll1$, Eqn.(\ref{eqn:10}) is essentially true only for negative values of the chemical potential, but not conclusive for positive values of the chemical potential.

On the other hand, for the case of very strong magnetic field ($b\gg1$), we can use the expansion:
\begin{eqnarray}\label{eqn:11}
\coth(jb)=\frac{\text{e}^{jb}+\text{e}^{-jb}}{\text{e}^{jb}-\text{e}^{-jb}}=1+2\sum_{k=1}^\infty\text{e}^{-2kjb},
\end{eqnarray}
and to the leading order in $b$, we approximate Eqn.(\ref{eqn:9}) as
\begin{eqnarray}\label{eqn:12}
\delta\Omega\approx-\frac{Nk_BT}{\text{Li}_{3/2}(-z)}\bigg[\frac{\mu_BB}{k_BT}\text{Li}_{3/2}(-z)-\text{Li}_{5/2}(-z)\bigg].
\end{eqnarray}
This equation leads to the saturation of magnetization of our system.

Eqns.(\ref{eqn:10}) and (\ref{eqn:12}) do not, of course, describe the magnetism of our system for all temperatures and fields, and are not true in general except in the two extreme cases. However, the expansion in Eqn.(\ref{eqn:11}) is true for all values of $b$. Using this expansion and summing over $j$, we recast Eqn.(\ref{eqn:9}) as
\begin{eqnarray}\label{eqn:13}
\delta\Omega&=&-\frac{Nk_BT}{\text{Li}_{\frac{3}{2}}(-z)}\bigg[2b\big[\frac{1}{2}\text{Li}_{\frac{3}{2}}(-z)+\sum_{k=1}^\infty\text{Li}_{\frac{3}{2}}(-\text{e}^{\nu-2kb})\big]\nonumber\\&&-\text{Li}_{\frac{5}{2}}(-z)\bigg],
\end{eqnarray}
where $\nu=\mu/k_BT$ is a scaled chemical potential. Eqn.(\ref{eqn:13}) is also the most general equation like Eqn.(\ref{eqn:9}) to describe the magnetism of our system. It is a novel form, and is easier to work with in comparison to the form in Eqn.(\ref{eqn:9}). In the next, we will evaluate the magnetization of our system for all values of the magnetic field as well as for all values temperature using the form obtained in Eqn.(\ref{eqn:13}).

\subsection{Asymptotic expansion of the grand potential}
Sommerfeld's asymptotic expansion of Fermi integral (polylog function) is of common interest for the low temperature properties of Fermi systems \cite{huang}. But, for the system of our interest, this expansion does not work well in particular for the oscillatory part of the grand potential in Eqn.(\ref{eqn:13}). It works well only for $\nu\rightarrow\infty$ resulting
\begin{eqnarray}\label{eqn:13i}
\delta\Omega&\approx&-N\mu_BB\bigg[1-\frac{2\mu}{5\mu_BB}-4\sqrt{2}\big(\frac{\mu_BB}{\mu}\big)^{3/2}\nonumber\\&&\times\text{Im}\big[\zeta(-3/2,1-\frac{\mu}{2\mu_BB})-\zeta(-3/2)\big]\bigg],
\end{eqnarray}
where $\zeta(s,a)=\sum_{n=0}^{\infty}1/(a+n)^s$ is a Hurwitz Zeta function. It is easy to check that, this asymptotic result differs significantly from the exact result even for $\nu\gg1$. In the following, we will present a novel asymptotic expansion technique for the grand potential adopting Landau's calculation for the de Haas-van Alphen effect \cite{landau,landau-lifshitz}.

It is nice to look at the $b$ dependent part at the right hand side of Eqn.(\ref{eqn:13}). It appears as the same form as in the left hand side of Eqn.(\ref{eqn:5}) which, of course, is a different form of Poisson summation formula. Using this summation formula in Eqn.(\ref{eqn:13}) and evaluating the integrations once by parts we get
\begin{eqnarray}\label{eqn:13a}
\delta\Omega=\frac{8Nk_BTb^2}{-\text{Li}_{\frac{3}{2}}(-z)}\sum_{k=1}^\infty\int_0^\infty\text{Li}_{\frac{1}{2}}(-\text{e}^{\nu-2xb})\frac{\sin(2\pi kx)}{2\pi k}\text{d}x.~~
\end{eqnarray}
Replacing the polylog function in Eqn.(\ref{eqn:13a}) by its integral from, we get
\begin{eqnarray}\label{eqn:13b}
\delta\Omega&=&\text{Im}\bigg[\frac{8Nk_BTb^2}{\text{Li}_{3/2}(-z)\Gamma(1/2)}\sum_{k=1}^\infty\frac{1}{2\pi k}\int_0^\infty\int_0^\infty\nonumber\\&&\frac{y^{-1/2}\text{e}^{i2\pi kx}}{1+\text{e}^{y-(\nu-2xb)}}\text{d}y\text{d}x\bigg].
\end{eqnarray}
Now, replacing the integration variable $x$ by $z=2bx+y-\nu$ and integrating over $z$ we recast Eqn.(\ref{eqn:13b}) as
\begin{eqnarray}\label{eqn:13c}
\delta\Omega&=&\text{Im}\bigg[\frac{4Nk_BTb}{\text{Li}_{\frac{3}{2}}(-z)\sqrt{\pi}}\sum_{k=1}^\infty\frac{1}{2\pi k}\int_0^\infty\big[-\text{e}^{\frac{k\pi^2}{b}}\text{B}(-\text{e}^{y-\nu},\nonumber\\&&\frac{ik\pi}{b},0)-i\pi\text{csch}(\frac{k\pi^2}{b})\big]\text{e}^{i\pi k(\nu-y)/b}y^{-1/2}\text{d}y\bigg],
\end{eqnarray}
where $\text{B}(-\text{e}^{y-\nu},\frac{ik\pi}{b},0)$ is an incomplete beta function, and for large $\nu$, its asymptotic series expansion is given by
\begin{eqnarray}\label{eqn:13d}
\text{e}^{\frac{k\pi^2}{b}}\text{B}(-\text{e}^{y-\nu},\frac{ik\pi}{b},0)=\text{e}^{\frac{i\pi k(y-\nu)}{b}}\sum_{j=0}^\infty\frac{(-1)^{j}b\text{e}^{j(y-\nu)}}{jb+i\pi k}.~~
\end{eqnarray}
Using this expansion formula and evaluating the integrations for $0\le y\le\nu$ \& $\nu\gg1$, we recast Eqn.(\ref{eqn:13c}) as
\begin{eqnarray}\label{eqn:13e}
\delta\Omega&\approx&\text{Im}\bigg[\frac{4Nk_BTb}{\frac{-\nu^{3/2}\sqrt{\pi}}{\Gamma(5/2)}}\sum_{k=1}^\infty\frac{1}{2\pi k}\bigg\{\frac{2ib\sqrt{\nu}}{\pi k}-\text{e}^{i(\frac{\pi k\nu}{b}+\frac{\pi}{4})}\pi\times\nonumber\\&&\sqrt{\frac{b}{k}}\text{csch}(\frac{k\pi^2}{b})+\sum_{j=1}^\infty\frac{(-1)^{j+1}2bF(\sqrt{j\nu})}{(jb+i\pi k)\sqrt{j}}\bigg\}\bigg],
\end{eqnarray}
where $F(x)=\text{e}^{-x^2}\int_0^x\text{e}^{y^2}\text{d}y$ is a Dawson integral. Our result in Eqn.(\ref{eqn:13e}) is not only a generalization of Pauli-Landau result, but also is an exact formula for $T\rightarrow0$. First two terms of this equation lead to Pauli-Landau (or Sondheimer-Wilson) formula \cite{landau-lifshitz,landau,sondheimer,wilson}
\begin{eqnarray}\label{eqn:13f}
\delta\Omega_{P-L}&=&-\frac{N}{2}\bigg[\frac{(\mu_BB)^2}{\mu}-3k_BT\bigg(\frac{\mu_BB}{\mu}\bigg)^{3/2}\times\nonumber\\&&\sum_{k=1}^\infty\frac{\text{csch}(\frac{k\pi^2k_BT}{\mu_BB})}{k^{3/2}}\cos\big(\frac{k\pi\mu}{\mu_BB}-\frac{\pi}{4}\big)\bigg].
\end{eqnarray}
The second term in Eqn.(\ref{eqn:13e}) as well as in Eqn.(\ref{eqn:13f}), represents de Haas-van Alphen oscillations around the contribution of the first term which represents the combination of Pauli paramagnetism and Landau diamagnetism. It is clear from Eqn.(\ref{eqn:13f}) that, the oscillatory term plays a significant role only in the strong field regime ($k_BT\lesssim\mu_BB\ll\mu$). It dies out exponentially in the weak field regime ($\frac{\mu_BB}{k_BT}\ll1$). This oscillatory term actually was obtained by Landau in 1939 \cite{landau,landau-lifshitz}. This term forms the basis of the de Haas-van Alphen effect. In the low temperature regime ($k_BT\ll\mu$), Eqn.(\ref{eqn:13f}) works well for weak and strong field cases, but not for very strong field case ($k_BT\ll\mu\lesssim\mu_BB$). On the other hand, our approximate result in Eqn.(\ref{eqn:13e}) works well for all strengths of magnetic field in particular for the low temperature, and is capable of predicting saturation of magnetization for very strong field case. Eqn.(\ref{eqn:13e}) represent a novel asymptotic series expansion for the grand potential of the 3-D gas of electrons exposed in a constant magnetic field. Low temperature properties of this system can now be easily obtained from this expression. In the following, we will compare Pauli-Landau result with that of ours evaluating magnetization of the ideal gas of electrons.

\section{Evaluation of magnetization and susceptibility}
Magnetization ($M\hat{k}$) of our system is to be obtained using the definition $\textbf{M}=-\frac{1}{V}\frac{\partial\delta\Omega}{\partial\textbf{B}}$. Similarly, Pauli-Landau result on the magnetization ($M_{P-L}\hat{k}$) is to be obtained from Eqn.(\ref{eqn:13f}) as \cite{landau-lifshitz}
\begin{eqnarray}\label{eqn:13g}
M_{P-L}(T,B)=-\frac{1}{V}\frac{\partial\delta\Omega_{P-L}}{\partial B}.
\end{eqnarray}
Right side of Eqn.(\ref{eqn:13g}) has already been plotted in FIG. 1 using an almost exact temperature dependent formula of the chemical potential \cite{biswas}. We will illustrate the temperature dependence of the chemical potential later in this section. Before that, let us see how the exact form of the magnetization looks like.

\subsection{Temperature and field dependence of the magnetization}
Now, we obtain expression of the magnetization (taking the partial derivative of $\delta\Omega$ in Eqn.(\ref{eqn:13})) as
\begin{eqnarray}\label{eqn:14}
M(T,B)&=&\frac{\bar{n}\mu_B}{\text{Li}_{\frac{3}{2}}(-z)}\bigg[2\big[\frac{1}{2}\text{Li}_{\frac{3}{2}}(-z)+\sum_{k=1}^\infty\text{Li}_{\frac{3}{2}}(-\text{e}^{\nu-2kb})\big]\nonumber\\&&-2b\sum_{k=1}^\infty2k\text{Li}_{\frac{1}{2}}(-\text{e}^{\nu-2kb})\bigg],
\end{eqnarray}
where $\bar{n}=N/V$ is the fixed average number density of particles (electrons). It is clear from Eqn.(\ref{eqn:14}) that, as $b\rightarrow\infty$, only the first term in the square bracket contributes to yield saturation value of the magnetization as $M_s=\bar{n}\mu_B$. We recast Eqn.(\ref{eqn:14}) with this saturation value as
\begin{eqnarray}\label{eqn:15}
M(T,B)&=&M_s\bigg[1+\frac{2}{\text{Li}_{\frac{3}{2}}(-\text{e}^{\nu})}\sum_{k=1}^\infty\big\{\text{Li}_{\frac{3}{2}}(-\text{e}^{\nu-2kb})\nonumber\\&&-2bk\text{Li}_{\frac{1}{2}}(-\text{e}^{\nu-2kb})\big\}\bigg],
\end{eqnarray}
This is the most general formula for the magnetization of our system. For $b\ll1$, using Euler-Maclaurin summation formula in Eqn.(\ref{eqn:15}), we approximate $M$ as
\begin{eqnarray}\label{eqn:16}
M(T,B)&\approx&\frac{M_s}{\text{Li}_{3/2}(-z)}\bigg[\frac{2b}{3}\text{Li}_{1/2}(-z)-\frac{4b^3}{45}\text{Li}_{-3/2}(-z)\nonumber\\&&+\frac{4b^5}{315}\text{Li}_{-7/2}(-z)-{\it{O}}(b^7)\bigg].
\end{eqnarray}
Although this approximate formula, to the first order in $b$, gives rise to the unification of Pauli paramagnetism and Landau diamagnetism \cite{landau-lifshitz,huang,wilson}:
\begin{eqnarray}\label{eqn:17}
M_{\text{p+d}}(T,B)=\frac{2M_s}{3}\frac{\mu_BB}{k_BT}\frac{\text{Li}_{1/2}(-z)}{\text{Li}_{3/2}(-z)},
\end{eqnarray}
the third and higher order terms in $b$ individually go to infinite for $z\ge1$ with alternative signs. Thus, it is difficult to judge whether $M_{\text{p+d}}$ be the leading term for $b\ll1$ until we compare $M_{\text{p+d}}$ and $M$ plotting in a figure. To plot them with respect to the magnetic field keeping temperature fixed, we need to know the temperature dependence of the fugacity ($z$) or of the chemical potential ($\mu$).

\subsection{Temperature dependence of the chemical potential}
Chemical potential of our system by no means can be obtained in an exact temperature dependent formula. But, an almost exact formula for the same was previously obtained (within an approximation technique) using Eqn.(\ref{eqn:8}) as \cite{biswas}
\begin{eqnarray}\label{eqn:18}
\mu(t)\approx\mu_{\lnsim}(t)\theta(0.723-t)+\mu_{\gtrsim}(t)\theta(t-0.723),
\end{eqnarray}
where $t=T/T_F$ is our new scaled temperature, $T_F=\frac{\hbar^2}{2mk_B}(3\pi^2\bar{n})^{2/3}$ is the Fermi temperature of our system in absence of the magnetic field, $\theta(t)$ is a unit step function,
\begin{eqnarray}\label{eqn:19}
\frac{\mu_{\lnsim}(t)}{k_BT_F}\approx\bigg[1-\frac{\pi^2}{12}t^2-\frac{\pi^4}{80}t^4+{\it{O}}(t^6)\bigg]
\end{eqnarray}
is the chemical potential (in units of $k_BT_F$) in the quantum regime,
\begin{eqnarray}\label{eqn:20}
\frac{\mu_{\gtrsim}(t)}{k_BT_F}&\approx& t\ln\bigg[\sqrt{\frac{3}{8}}-\frac{(432\sqrt{3}-162)\pi^{\frac{1}{2}}t^{\frac{3}{2}}}{36\cdot2^{\frac{1}{3}}\cdot3^{\frac{5}{6}}f_3(t)}\nonumber\\&&+\frac{t^{-\frac{3}{2}}f_3(t)}{2\pi^{\frac{1}{2}}\cdot2^{\frac{2}{3}}\cdot3^{\frac{1}{6}}}\bigg]
\end{eqnarray}
is the chemical potential  (in units of $k_BT_F$) in the semi classical regime, and $f_3(t)=\big[192\pi t^3-[36\sqrt{6}-9\sqrt{2}]\pi^{\frac{3}{2}}t^{9/2}+\sqrt{3}\big(12288\pi^2t^6+[1152\sqrt{2}-4608\sqrt{6}]\pi^{5/2}t^{15/2}-[864-3072\sqrt{3}]\pi^3t^9\big)^{1/2}\big]^{1/3}$.

It should be mentioned, that, $\mu=\mu(t)$ is the actual chemical potential of our system, as because, it is treated as independent of the magnetic field absorbing the field dependence into the single particle energy levels \cite{landau-lifshitz,landau}.

\subsection{Plotting of magnetization and comparison with Pauli-Landau result}
\begin{figure}
\includegraphics{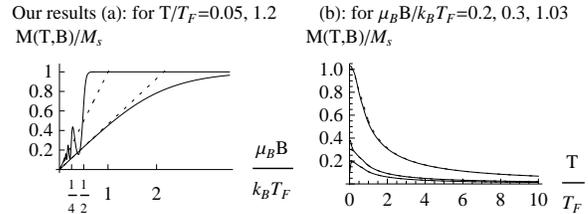}
\caption{Thick and thin solid lines in (a) represent our results (Eqn.(\ref{eqn:15})) for $T/T_F=0.05$ and $1.2$ respectively. Thick, semithick, and thin solid lines in (b) represent Eqn.(\ref{eqn:15}) for $\mu_BB/k_BT_F=0.2$, $0.3$ and $1.03$ respectively. Dashed and dotted lines represent respective weak field results (Eqn.(\ref{eqn:17})).}
\end{figure}

We plot the right hand side of Eqn.(\ref{eqn:15}) in FIG. 2 introducing above temperature dependence of the chemical potential in Eqn.(\ref{eqn:18}). Thick and thin solid lines in FIG 2 (a) represent our results (Eqn.(\ref{eqn:15})) on magnetic field dependence of the magnetization of the 3-D ideal gas of electrons for $T/T_F=0.05$ and $1.2$ respectively. The thick and thin dashed lines in this figure represent weak field results (Eqn.(\ref{eqn:17})) for $T/T_F=0.05$ and $1.2$ respectively. The oscillations of the magnetization around the respective dashed lines represent de Haas-van Alphen effect. The oscillations, according to Eqn.(\ref{eqn:15}) (and Eqn.(\ref{eqn:13f}) as well) are expected to be dying out exponentially as the chemical potential decreases with the increase of temperature. And, it is the case for the thin solid line. Thick, semithick, and thin solid lines in FIG. 2 (b) represent our results (Eqn.(\ref{eqn:15})) on temperature dependence of the magnetization for $\mu_BB/k_BT_F=0.2$, $0.3$ and $1.03$ respectively. Dotted lines in this figure represent corresponding weak field results.

For consistency, our result must have to match well with Pauli-Landau result in particular for the weak ($\mu_BB\ll k_BT$) and strong field cases ($k_BT\lesssim\mu_BB\ll k_BT_F$) at low temperatures ($T\ll T_F$). To show how Pauli-Landau result differs from that of ours in the very strong field regime ($\mu_BB\gtrsim k_BT_F$), we compare field dependence of both the results in FIG. 3 for a typical temperature, say $T/T_F=0.1$. From this figure one can conclude that, Pauli-Landau result is reasonably good even for room temperature and for weak \& strong magnetic field cases, but not for very strong magnetic field case. Unlike Pauli-Landau result, ours have the saturation limit as well as the classical limit \cite{singh,shaykhutdinov}.

While the magnetization approaches the classical limit for $T\gnsim T_F$, it oscillates around the weak field limit and approaches the saturation limit for $T\lesssim T_F$. But, these oscillations are not clearly apparent in FIGs. 2 and 3 in particular for the lower magnetic field regime. To make it apparent, we can plot its derivative (susceptibility).

\begin{figure}
\includegraphics{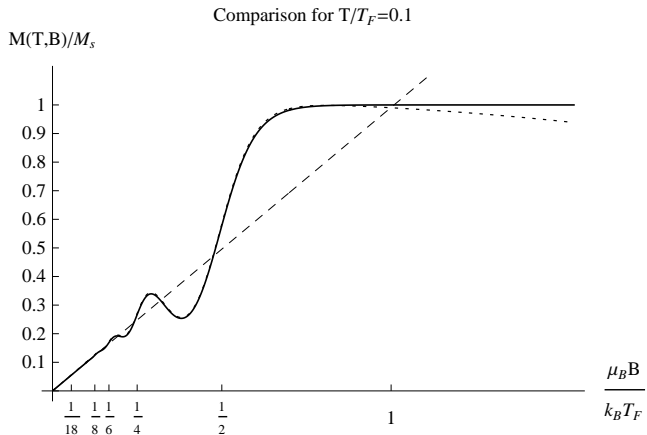}
\caption{Solid (our result), dotted (Pauli-Landau result) and dashed (weak field result) lines represent Eqn.(\ref{eqn:15}), Eqn.(\ref{eqn:13g}), and Eqn.(\ref{eqn:17}) for $T=0.1T_F$ respectively.}
\end{figure}

\subsection{Susceptibility}
Susceptibility ($\chi$) of our system is to be obtained using the definition $\chi(T,B)=\mu_0\frac{\partial M(T,B)}{\partial B}$. Now, from Eqn.(\ref{eqn:15}), we get the exact form of the susceptibility of our system as
\begin{eqnarray}\label{eqn:21}
\chi(T,B)&=&\mu_0\frac{\bar{n}\mu_B}{k_BT}\bigg[\frac{8}{\text{Li}_{\frac{3}{2}}(-\text{e}^{\nu})}\sum_{k=1}^\infty\big\{bk^2\text{Li}_{-\frac{1}{2}}(-\text{e}^{\nu-2kb})\nonumber\\&&-k\text{Li}_{\frac{1}{2}}(-\text{e}^{\nu-2kb})\big\}\bigg].
\end{eqnarray}
This formula of susceptibility is valid for all strengths of the magnetic field as well as for all temperatures. On the other hand, weak field result for the susceptibility ($\chi_{\text{p+d}}(T,B)=\mu_0\frac{\partial M_{\text{p+d}}(T,B)}{\partial B}$) is obtained from Eqn.(\ref{eqn:17}) as
\begin{eqnarray}\label{eqn:22}
\chi_{\text{p+d}}(T,B)=\mu_0\frac{2\bar{n}\mu_B^2}{3k_BT}\frac{\text{Li}_{1/2}(-z)}{\text{Li}_{3/2}(-z)}.
\end{eqnarray}
We plot the right hand sides of Eqns.(\ref{eqn:21}) in FIG. 4 introducing the temperature dependence of the chemical potential in Eqn.(\ref{eqn:18}). Thick and thin solid lines in FIG. 4 (a) represent our results (Eqn.(\ref{eqn:21})) on magnetic field dependence of the susceptibility for $T/T_F=0.1$ and $1.2$ respectively. Thick and thin dotted lines in this figure represents weak field limits (Eqn.(\ref{eqn:22})) for $T/T_F=0.1$ and $1.2$ respectively. The oscillations of the susceptibility around the respective dotted lines also represent de Haas-van Alphen effect. Thick and thin solid lines in FIG. 4 (b) represent our results (Eqn.(\ref{eqn:21})) on temperature dependence of the susceptibility for $\mu_BB/k_BT_F=0.2$ and $1.03$ respectively. Dotted line in this figure represents the weak field limit (Eqn.(\ref{eqn:22})). It is easy to check that, for $T\rightarrow0$, the susceptibility does not only oscillates rapidly with $1/B$, but also diverges as $B^{-3/2}$ for $B\rightarrow0$.

For the oscillatory nature of the magnetization, the susceptibility may have negative values which are evident in FIG. 4. So, a natural question arises, whether the specific heat of our system may have negative values. To know the same we have to evaluate temperature dependence of the thermodynamic energy of our system for different values of the magnetic field.

\begin{figure}
\includegraphics{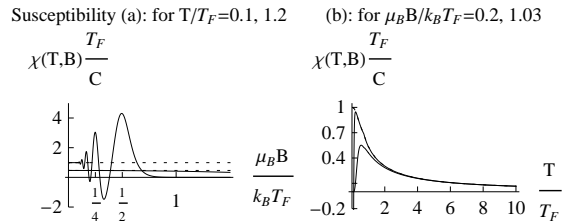}
\caption{Thick and thin solid lines in (a) represent our results (Eqn.(\ref{eqn:21})) for $T/T_F=0.1$ and $1.2$ respectively. Thick and thin solid lines in (b) represent Eqn.(\ref{eqn:21}) for $\mu_BB/k_BT_F=0.2$ and $1.03$ respectively. The dotted lines  represent respective weak field results (Eqn.(\ref{eqn:22})). Here $\text{C}=\mu_0\bar{n}\mu_B^2/k_B$ is the Curie constant.}
\end{figure}

\section{Thermodynamic energy}
Although low temperature specific heat of the system of our interest was calculated even for the interacting case \cite{luttinger,holstein}, yet the specific heat for the ideal case, has not surprisingly been calculated for all temperatures and fields. In the following we will also not calculate the specific heat, but will calculate the thermodynamic energy of the system for all temperatures as well as for all strengths of the magnetic field. Specific heat can, of course, be easily apparent from the thermodynamic energy.

If we do not want to restrict the average number of particles in Eqn.(\ref{eqn:13}) then, this equation can be recast along with Eqn.(\ref{eqn:8}) as
\begin{eqnarray}\label{eqn:23}
\delta\Omega&=&\frac{2Vk_BT}{\lambda_T^3}\bigg[\frac{2\mu_BB}{k_BT}\big[\frac{1}{2}\text{Li}_{\frac{3}{2}}(-z)+\sum_{k=1}^\infty\text{Li}_{\frac{3}{2}}(-\text{e}^{\frac{\mu-2k\mu_BB}{k_BT}})\big]\nonumber\\&&-\text{Li}_{\frac{5}{2}}(-z)\bigg].
\end{eqnarray}
Using the thermodynamic relation: $\delta E=\delta\Omega-T\frac{\partial\delta\Omega}{\partial T}|_{\mu,V}-\mu\frac{\partial\delta\Omega}{\partial\mu}|_{T,V}$ (for a grand canonical ensemble), we get magnetic field dependent part of the thermodynamic energy as
\begin{eqnarray}\label{eqn:24}
\delta E&=&-\frac{2Vk_BT}{\lambda_T^3}\bigg[\big\{b\big[\text{Li}_{\frac{3}{2}}(-z)+2\sum_{k=1}^\infty\text{Li}_{\frac{3}{2}}(-\text{e}^{\nu-2kb})\big]\nonumber\\&&-3\text{Li}_{\frac{5}{2}}(-z)\big\}+4b^2\sum_{k=1}^\infty k\text{Li}_{\frac{1}{2}}(-\text{e}^{\nu-2kb})\bigg].
\end{eqnarray}
Now, the field dependent part of the thermodynamic energy per particle ($\delta U=\delta E/N$) is given by
\begin{eqnarray}\label{eqn:25}
\delta U&=&\frac{k_BT}{\text{Li}_{\frac{3}{2}}(-z)}\bigg[\big\{b\big[\text{Li}_{\frac{3}{2}}(-z)+2\sum_{k=1}^\infty\text{Li}_{\frac{3}{2}}(-\text{e}^{\nu-2kb})\big]\nonumber\\&&-3\text{Li}_{\frac{5}{2}}(-z)\big\}+4b^2\sum_{k=1}^\infty k\text{Li}_{\frac{1}{2}}(-\text{e}^{\nu-2kb})\bigg].
\end{eqnarray}
Weak field expansion of the above can be obtained from the Euler-Maclaurin summation formula, and it results
\begin{eqnarray}\label{eqn:26}
\delta U&\approx&\frac{k_BT}{\text{Li}_{\frac{3}{2}}(-z)}\bigg[-\frac{b^2}{3}\text{Li}_{\frac{1}{2}}(-z)+\frac{b^4}{9}\text{Li}_{-\frac{3}{2}}(-z)\nonumber\\&&-\frac{2b^6}{105}\text{Li}_{-\frac{7}{2}}(-z)+{\it{O}}(b^8)\bigg].
\end{eqnarray}

\begin{figure}
\includegraphics{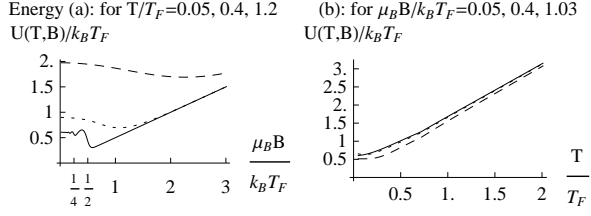}
\caption{Solid, dotted, and dashed lines in (a) represent our results (Eqn.(\ref{eqn:28})) for $T/T_F=0.05$, $0.4$ and $1.2$ respectively. Solid, dotted, and dashed lines in (b) represent Eqn.(\ref{eqn:28}) for $\mu_BB/k_BT_F=0.05$, $0.4$ and $1.03$ respectively.}
\end{figure}

It is interesting to note that, field dependent part of the total energy is the same as the field dependent part of the grand potential only in the weak field regime. Field independent part of the energy per particle, on the other hand, is given by \cite{huang,landau-lifshitz,biswas}
\begin{eqnarray}\label{eqn:27}
U_0=\frac{3}{2}k_BT\frac{\text{Li}_{\frac{5}{2}}(-z)}{\text{Li}_{\frac{3}{2}}(-z)}.
\end{eqnarray}
Thus, total thermodynamic energy per particle of our system is given by
\begin{eqnarray}\label{eqn:28}
U=U_0+\delta U.
\end{eqnarray}
We plot right hand side of Eqn.(\ref{eqn:28}) in FIG. 5 to see the magnetic field as well as temperature dependence of the thermodynamic energy of the system. Although the energy has oscillations with respect to the magnetic field, it never decreases with the increase of temperature. Thus, specific heat of our system, unlike the susceptibility, never gets a negative sign. It is also interesting to note that, energy weakly decreases in the weak field regime, oscillates with frequency $\sim1/B$ in the strong field regime, and increases linearly in very strong field regime. The last part of the interesting phenomena is a result of confinement of orbital motion of the electrons. It will not happen for spin magnets.

\section{Conclusions}
Pauli-Landau theory for the magnetism of the 3-D ideal gas of electrons is very much well established within the range of its applicability. But, it neither explains the saturation of magnetization nor it does have a classical limit. So, it is not a complete theory. Goal of this paper is to give a complete theory for the same. Our theory unifies Pauli paramagnetism, Landau diamagnetism, and de Haas-van Alphen effect for the 3-D ideal gas of electrons in a single framework. It is applicable for all temperatures as well as for all strengths of the magnetic field.

Our theory is consistent with the unified effect of Pauli paramagnetism, Landau diamagnetism and de Haas-van Alphen effect for weak and strong field cases. Our result goes beyond Pauli-Landau result, and is explicitly able to describe the range of the de Haas-van Alphen oscillations within the entire range of the magnetic field. Our result is also able to describe quantum to classical crossover as the temperature is increased.

While Pauli-Landau result consists of oscillatory terms over the weak field limit, our result consists of nonoscillatory terms of alternative signs over the saturation limit. We also have done a novel asymptotic analysis of our result, in particular, for the low temperature regime. This asymptotic analysis gives rise to a generalization of Pauli-Landau result, and it eases comparison of Pauli-Landau result with that of ours. Low temperature properties of the system can now be easily explored from our asymptotic analysis. For a given strength of the magnetic field, applicability of Pauli-Landau (or Sondheimer-Wilson) result can now be easily tested comparing with our result.

The last section (thermodynamic energy) has been added to explore the thermodynamics of the ideal gas of electrons in presence of a constant magnetic field. One can now easily obtain specific heat, entropy, equation of state, etc from the expression of the energy per particle.

In our whole analyses we have considered the chemical potential of the system to be independent of the magnetic field absorbing the field dependence entirely into the single particle energy levels \cite{landau-lifshitz,landau}. On the other hand, absorbing the field dependence even partially into the chemical potential, one may get de Haas-van Alphen type of oscillations of the chemical potential \cite{lukyanchuk,nizhankovskii}.

For electrons in a metal, Landau's formula (for the de Haas-van Alphen effect) gets a generalization to Lifshitz-Kosevich formula for inclusion of the effective mass of an electron within the lattice structure of the ions in the metal \cite{lifshitz,landau-lifshitz2}. Lifshitz-Kosevich formula is very much useful for the analysis of experimental data \cite{cooper,gasparov,bergk} even for the real (interacting Bloch) electrons in a metal with further modification due to interparticle interactions \cite{luttinger2}. But, like Landau's formula, Lifshitz-Kosevich formula is also not applicable for all temperatures and fields. Generalizing our result, for the interacting Bloch electrons in a metal, one can complete Lifshitz-Kosevich formula, and can apply the generalized complete formula for all temperatures and fields.

\section*{Acknowledgment}
This work has been sponsored by the University Grants Commission [UGC] under the D.S. Kothari Postdoctoral Fellowship Scheme {[No.F.4-2/2006(BSR)/13-280/2008(BSR)]}.

\end{document}